\documentclass[twocolumn,showpacs,preprintnumbers,amsmath,amssymb]{revtex4}
\usepackage{graphicx}
\usepackage{amsmath}

\begin{document}
\title{Geometry of depolarizing channels}
\author{Kuldeep Dixit}
\email{kuldeep@physics.utexas.edu}
\author{E. C. G. Sudarshan}
\affiliation{Department of Physics,
        University of Texas,
        Austin, TX 78712.}

\begin{abstract}
Depolarizing maps acting on an $N$ dimensional system are
completely positive maps resulting into compression of the Bloch
`ball' along $N^2 -1$ polarization directions. In the qubit case
these maps are a convex sum of four extremal maps and form a
simplex in the space of compression coefficients along the three
polarization directions. We calculate the compression domain for
three and four level systems. For a three level system the region
has curved surfaces, but it is a simplex for a four level system.
We conjecture that it is a simplex in the case of $2^{n}$ level
systems.
\end{abstract}

\date{\today} \maketitle

\section{Introduction}
The dynamics of open quantum systems can be described in terms of
Completely Positive (CP), trace preserving maps acting on the
system \cite{3,4,5,6,7,8}. In general these maps can rotate,
compress and translate the Bloch ball, resulting into `Affine'
maps \cite{13}. These maps are used to simulate different noises
that could be acting on the system \cite{1,2}.

A simple example of noise can be given by depolarizing maps, which
result into shrinking of the Bloch ball. In general the
compression can be anisotropic, giving a rich geometric structure
to the problem. The requirement of complete positivity turns out
to be stronger than just positivity. For example, for N=2 the map
corresponding to shrinking of the Bloch sphere only along one
direction and leaving the other two invariant (pancake map) is not
CP. In N=2 case the general depolarizing map has a very simple
structure and can be written as a convex sum of four fixed
extremal maps. One would like to see if it true for higher
dimensional systems.

In what follows we answer the question for N=3 and N=4 cases. We
also give a way to answer the question for higher level systems,
although we must admit that the calculations becomes much more
involved.

Our paper is organized as follows: In Sec. II we briefly describe
the geometry of N=2 level depolarizing channels. In Sec. III we
show how to calculate the region of positivity for an N
dimensional system. We carry out the calculations for N=3 and N=4
cases in Sec. IV and V respectively. We conclude the paper with
further discussions in Sec. VI.

\section{The qubit case} The qubit density matrix is given by

\begin{equation}
\rho=\frac{1}{2}\left(I + a\cdot \sigma \right).
\end{equation}
Positivity of $\rho$ requires that $\overrightarrow{a}$ lies
inside a sphere with unit radius. Let's assume that due to
depolarization, the sphere shrinks along the three polarization
directions by factors of $\nu_1$, $\nu_2$ and $\nu_3$. The natural
question arises - what are the allowed values of $\nu_1$, $\nu_2$
and $\nu_3$ to ensure complete positivity of the map?

It turns out \cite{12} that the allowed region in the space of
$\nu_1$, $\nu_2$ and $\nu_3$ forms a tetrahedron with vertices at
(1,1,1), (1,-1,-1), (-1,1,-1) and (-1,-1,1). These vertices correspond to four extremal maps given by
$B_i(\rho)=\sigma_i \rho \sigma_i$, where $i=0,1,2,3$ and
$\sigma_0=I_2$. The general depolarizing map can be written as a
convex sum of these maps as
\begin{equation}
B(\rho)=p \sigma_1 \rho \sigma_1 + q\sigma_2 \rho \sigma_2+
r\sigma_3 \rho \sigma_3 +(1-p-q-r) \rho
\end{equation}
where p, q, r are non-negative.

Only the region inside the tetrahedron is CP. For example, partial
transpose corresponds to the point (1,-1,1) and lies outside the
region.

\begin{figure}[htp]
\centering
\includegraphics[width=6 cm]{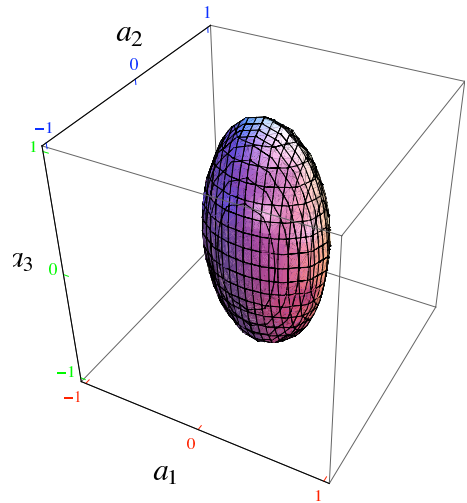}
\caption{As a result of depolarization the Bloch sphere shrinks
anisotropically}\label{fig:halla}
\end{figure}

\begin{figure}[htp]
\centering
\includegraphics[width=6 cm]{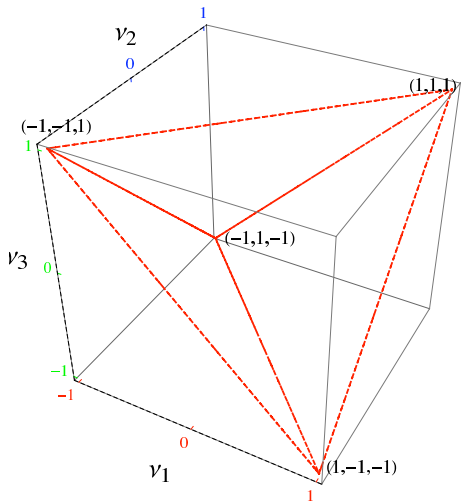}
\caption{CP region in the space of $\nu$'s is a tetrahedron with
vertices corresponding to maps $1 \rho 1$, $\sigma_1 \rho
\sigma_1$, $\sigma_2 \rho \sigma_2$ and $\sigma_3 \rho \sigma_3$}
\label{fig:gulla}
\end{figure}

\section{The N level system}
For an N level system the density matrix can be written in terms
of $N^2-1$ traceless, Hermitian and
 trace-orthogonal generators of SU(N), denoted by $J_i$, and the
 polarizations
 along those directions ($a_i$) as
\begin{equation}
\rho=\frac{1}{N}\left(I_n+\sqrt{\frac{n(n-1)}{2}}\sum_{i=1}^{N^2-1}
a_i J_i\right).
\end{equation}
This normalization results into pure states having unit length
\cite{10}, analogous to the qubit case.

Let us assume that a general map acting on $\rho$ takes it into
another density matrix given by $\rho'$. The map can be written in
terms of the $A$ matrix\cite{3} as:

\begin{equation}
\rho'_{rs}=A_{rs,r's'}\rho_{r's'}.
\end{equation}
With change of indices, we can find another matrix B which is
Hermitian and has simpler conditions of complete positivity. It is
obtained by
\begin{equation}
B_{rr',ss'}=A_{rs,r's'}.
\end{equation}

Complete positivity requires B to be positive semi-definite (and
thus have non-negative eigenvalues).

\subsection{Deriving A from compression of the
coherence vector}
  The simplest way to understand the depolarizing map is to think of it as
compression of the coherence vector. Let's assume that as the
result of the map the polarizations are modified as
\begin{equation}a'_i= \nu_i a_i .
\end{equation}
which can be considered as a special case of
\begin{equation}
\label{eq:diag} a'_i= T_{ij} a_j =(\delta_{ij} \nu_j) a_j.
\end{equation}
For completeness we take T to be $N^2\otimes N^2$ dimensional
matrix acting on the vector (1,$\overrightarrow{a}$). Needless to
say, $T_{11}=1$ as required by the trace preserving condition.
Both the T and the A matrices contain all the information about
the map and are thus related. The relation can be established in
the following way.

 Since the super-matrix A acts on all the entries in $\rho$, we can convert
$\rho$ into a vector (let's call it $V^{\rho}$) which is acted
upon by A. The entries in $V^{\rho}$ are linearly related to the
coherence vector as
\begin{equation} V^{\rho}_i=M_{ij}a_j
\end{equation}
The map takes $V^{\rho}$ to $V^{\rho'}$ which is related to
$V^{\rho}$ as
\begin{equation}
V^{\rho'}=M \cdot a'= M \cdot T \cdot a =A \cdot V^{\rho}=A \cdot
M \cdot a.
\end{equation}
Thus
\begin{equation} A=M \cdot T \cdot M^{-1}.
\end{equation}

Once we get A and B we can find the eigenvalues of B and check for
complete positivity. We solve the case of N=3 and N=4 in next two
sections.

\section{N=3 case}
The qutrit density matrix is written as
\begin{equation}
\rho=\frac{1}{3}(1+\sqrt{3}\sum_{i=1}^8 a_i \lambda_i)
\end{equation}
where $\lambda's$ are  the Gell-Mann matrices

\begin{equation}
\begin{array}{ccc}
  \lambda_1=\left(
\begin{array}{ccc}
 0 & 1 & 0 \\
 1 & 0 & 0 \\
 0 & 0 & 0
\end{array}
\right), & \lambda_2=\left(
\begin{array}{ccc}
 0 & -i & 0 \\
 i & 0 & 0 \\
 0 & 0 & 0
\end{array}
\right), & \lambda_3=\left(
\begin{array}{ccc}
 1 & 0 & 0 \\
 0 & -1 & 0 \\
 0 & 0 & 0
\end{array}
\right), \\
  \lambda_4=\left(
\begin{array}{ccc}
 0 & 0 & 1 \\
 0 & 0 & 0 \\
 1 & 0 & 0
\end{array}
\right), &  \lambda_5=\left(
\begin{array}{ccc}
 0 & 0 & -i \\
 0 & 0 & 0 \\
 i & 0 & 0
\end{array}
\right),& \lambda_6=\left(
\begin{array}{ccc}
 0 & 0 & 0 \\
 0 & 0 & 1 \\
 0 & 1 & 0
\end{array}
\right), \\
  \lambda_7=\left(
\begin{array}{ccc}
 0 & 0 & 0 \\
 0 & 0 & -i \\
 0 & i & 0
\end{array}
\right) , & \lambda_8=\frac{1}{\sqrt{3}}\left(
\begin{array}{ccc}
 1 & 0 & 0 \\
 0 & 1 & 0 \\
 0 & 0 & -2
\end{array}
\right). &  \\
\end{array}
\end{equation}

As shown in (\ref{eq:diag}), T can be written as
\begin{equation}
T=diag({1, \nu_1, \nu_2, ..., \nu_8})
\end{equation}

M relates the density matrix to the coherence vector as:

\begin{equation}
V^{\rho}_i=M_{ij}a_j.
\end{equation}

Explicitly:
\begin{widetext}
\begin{equation}
\frac{1}{3}
\left(%
\begin{array}{c}
  1+{\sqrt{3}}a_3+ a_8\\
 {\sqrt{3}} (a_1 -ia_2) \\
 {\sqrt{3}} (a_4 -ia_5 )\\
 {\sqrt{3}} (a_1 +ia_2 )\\
  1-{\sqrt{3}}a_3 +a_8 \\
  {\sqrt{3}}(a_6 -ia_7 )\\
  {\sqrt{3}}(a_4 +ia_5 ) \\
 {\sqrt{3}} (a_6 +ia_7  )\\
  1-2 a_8\\
\end{array}%
\right)=\frac{1}{3}\left(%
\begin{array}{ccccccccc}
  1&0&0&{\sqrt{3}}&0&0&0&0&1 \\
    0&{\sqrt{3}}&-{\sqrt{3}}i &0&0&0&0&0&0 \\
    0&0&0&0&{\sqrt{3}}&-{\sqrt{3}}i &0&0&0 \\
    0&{\sqrt{3}}&{\sqrt{3}}i &0&0&0&0&0&0 \\
    1&0&0&-{\sqrt{3}}&0&0&0&0&1 \\
    0&0&0&0&0&0&{\sqrt{3}}&-{\sqrt{3}}i &0 \\
    0&0&0&0&{\sqrt{3}}&{\sqrt{3}}i &0&0&0 \\
    0&0&0&0&0&0&{\sqrt{3}}&{\sqrt{3}}i &0 \\
    1&0&0&0&0&0&0&0&-2\\
\end{array}%
\right).\left(%
\begin{array}{c}
  1 \\
  a_1 \\
  a_2 \\
  a_3\\
  a_4\\
  a_5\\
  a_6 \\
  a_7 \\
  a_8 \\
\end{array}%
\right).
\end{equation}
\end{widetext}
Calculation for
$A=M \cdot T \cdot M^{-1}$
yields:
\begin{widetext}
\begin{equation}
A=\left(%
\begin{array}{ccccccccc}
\frac{1}{3}+\frac{\nu _3}{2}+\frac{\nu _8}{6} & 0 & 0 & 0 &
\frac{1}{3}-\frac{\nu _3}{2}+\frac{\nu _8}{6} & 0 & 0 & 0 &
\frac{1}{3}-\frac{\nu _8}{3}
\\
 0 & \frac{\nu _1}{2}+\frac{\nu _2}{2} & 0 & \frac{\nu _1}{2}-\frac{\nu _2}{2} & 0 & 0 & 0 & 0 & 0 \\
 0 & 0 & \frac{\nu _4}{2}+\frac{\nu _5}{2} & 0 & 0 & 0 & \frac{\nu _4}{2}-\frac{\nu _5}{2} & 0 & 0 \\
 0 & \frac{\nu _1}{2}-\frac{\nu _2}{2} & 0 & \frac{\nu _1}{2}+\frac{\nu _2}{2} & 0 & 0 & 0 & 0 & 0 \\
 \frac{1}{3}-\frac{\nu _3}{2}+\frac{\nu _8}{6} & 0 & 0 & 0 & \frac{1}{3}+\frac{\nu _3}{2}+\frac{\nu _8}{6} & 0 & 0 & 0 & \frac{1}{3}-\frac{\nu _8}{3}
\\
 0 & 0 & 0 & 0 & 0 & \frac{\nu _6}{2}+\frac{\nu _7}{2} & 0 & \frac{\nu _6}{2}-\frac{\nu _7}{2} & 0 \\
 0 & 0 & \frac{\nu _4}{2}-\frac{\nu _5}{2} & 0 & 0 & 0 & \frac{\nu _4}{2}+\frac{\nu _5}{2} & 0 & 0 \\
 0 & 0 & 0 & 0 & 0 & \frac{\nu _6}{2}-\frac{\nu _7}{2} & 0 & \frac{\nu _6}{2}+\frac{\nu _7}{2} & 0 \\
 \frac{1}{3}-\frac{\nu _8}{3} & 0 & 0 & 0 & \frac{1}{3}-\frac{\nu _8}{3} & 0 & 0 & 0 & \frac{1}{3}+\frac{2 \nu _8}{3}
\end{array}%
\right).
\end{equation}
\end{widetext}

The $B$ matrix is obtained by change of indices as

\begin{equation}
B_{rr',ss'}=A_{rs,r's'}
\end{equation}

and is given by:
\begin{widetext}
\begin{equation}
B=\left(%
\begin{array}{ccccccccc}
  \frac{2+{3\nu_3} + {\nu_8}}{6}&0&0&0&\frac{ {\nu_1}}{2}+\frac{ {\nu_2}}{2}&0&0&0&\frac{ {\nu_4 + \nu_5}}{2}
\\
    0&\frac{2-3\nu_3+{\nu_8}}{6}&0&\frac{ {\nu_1}}{2}-\frac{ {\nu_2}}{2}&0&0&0&0&0 \\
    0&0&\frac{1}{3}-\frac{ {\nu_8}}{3}&0&0&0&\frac{ {\nu_4}}{2}-\frac{ {\nu_5}}{2}&0&0 \\
    0&\frac{ {\nu_1}}{2}-\frac{ {\nu_2}}{2}&0&\frac{2-3\nu_3+{\nu_8}}{6}&0&0&0&0&0 \\
    \frac{ {\nu_1}}{2}+\frac{ {\nu_2}}{2}&0&0&0&\frac{2+3\nu_3+{\nu_8}}{6}&0&0&0&\frac{ {\nu_6 + \nu_7}}{2}
\\
    0&0&0&0&0&\frac{1}{3}-\frac{ {\nu_8}}{3}&0&\frac{ {\nu_6}}{2}-\frac{ {\nu_7}}{2}&0 \\
    0&0&\frac{ {\nu_4}}{2}-\frac{ {\nu_5}}{2}&0&0&0&\frac{1}{3}-\frac{ {\nu_8}}{3}&0&0 \\
    0&0&0&0&0&\frac{ {\nu_6}}{2}-\frac{ {\nu_7}}{2}&0&\frac{1}{3}-\frac{ {\nu_8}}{3}&0 \\
    \frac{ {\nu_4}}{2}+\frac{ {\nu_5}}{2}&0&0&0&\frac{ {\nu_6}}{2}+\frac{ {\nu_7}}{2}&0&0&0&\frac{1 +2 \nu_8}{3}
\\
\end{array}%
\right).
\end{equation}
\end{widetext}

For the map to be positive, all the eigenvalues should be positive
semi-definite. Six of these eigenvalues are given by the
`hyperplanes':

\begin{eqnarray}
h_1=\lambda_1 &=& \frac{1}{6}(2+3\nu_4-3\nu_5-2\nu_8) \nonumber \\
h_2=\lambda_2 &=& \frac{1}{6}(2-3\nu_4+3\nu_5-2\nu_8) \nonumber \\
h_3=\lambda_3 &=& \frac{1}{6}(2+3\nu_6-3\nu_7-2\nu_8) \nonumber \\
h_4=\lambda_4 &=& \frac{1}{6}(2-3\nu_6+3\nu_7-2\nu_8) \nonumber \\
h_5=\lambda_5 &=& \frac{1}{6}(2+3\nu_1-3\nu_2-3\nu_3+\nu_8) \nonumber \\
h_6=\lambda_6 &=& \frac{1}{6}(2-3\nu_1+3\nu_2-3\nu_3+\nu_8) \nonumber \\
\end{eqnarray}
The remaining three eigenvalues are given by the eigenvalues of
\begin{equation}
H=\left(%
\begin{array}{ccc}
  \frac{2+3\nu_3+{\nu_8}}{6} & \frac{\nu_1+\nu_2}{2} & \frac{\nu_4+\nu_5}{2} \\
  \frac{\nu_1+\nu_2}{2} & \frac{2+3\nu_3+{\nu_8}}{6} & \frac{\nu_6+\nu_7}{2} \\
  \frac{\nu_4+\nu_5}{2} & \frac{\nu_6+\nu_7}{2} & \frac{1+2\nu_8}{3} \\
\end{array}%
\right)
\end{equation}

Since the matrix is Hermitian and thus has real roots, the
condition of positive eigenvalues translates to
$(\lambda_7+\lambda_8+\lambda_9),
(\lambda_7\lambda_8+\lambda_8\lambda_9+\lambda_9\lambda_7),
\lambda_7\lambda_8\lambda_9$ being positive. These can be easily
found from the characteristic equation as

\begin{widetext}
\begin{eqnarray}
h_7&=&\lambda_7+\lambda_8+\lambda_9  \nonumber\\
   &=& 1+\nu_3+\nu_8 \\
 s_1&=&\lambda_7\lambda_8+\lambda_8\lambda_9+\lambda_9\lambda_7 \nonumber \\
    &=& \frac{1}{12} \left(-3 \left(\nu _1+\nu _2\right){}^2+3 \nu _3^2-3 \left(\nu _4+\nu _5\right){}^2-3 \left(\nu _6+\nu _7\right){}^2+\left(2+\nu _8\right) \left(2+3 \nu _8\right)+2 \nu _3 \left(4+5 \nu _8\right)\right). \\
s_2 &=& \lambda_7\lambda_8\lambda_9 \nonumber \\
    &=& \frac{1}{216}  (54 \nu _2  (\nu _4+\nu _5 )  (\nu _6+\nu
_7 )-18 \nu _1^2  (1+2 \nu _8 )-18 \nu _2^2
 (1+2 \nu _8 )+18 \nu _1  (3  (\nu _4+\nu
_5 )  (\nu _6+\nu _7 ) -2 \nu _2  (1+2 \nu _8 ) ) \nonumber \\
&+& (2+3 \nu _3+\nu _8 )  (4-9  (\nu _4+\nu _5 ){}^2-9  (\nu
_6+\nu _7 ){}^2+10 \nu _8+4 \nu _8^2+6 \nu _3  (1+2 \nu _8 ) ) ).
\end{eqnarray}
\end{widetext}

Hence we see that the completely positive region is bounded by
seven hyperplanes, a second degree surface $s_1$ and a third
degree surface $s_2$. This is quite different from the qubit case
where the allowed region is bounded by four planes and is just a
simplex with four vertices. For N=3 the depolarizing map has an
infinite number of extremal points corresponding to the curved
surface.

\section{N=4 case}
From the previous section we might expect
that the CP region will be a curved surface for N=4, but quite analogous
to the qubit case it turns out to be a simplex.

  Let's start with taking the generators to be:
$J$=\{$I\otimes \sigma_i$, $\sigma_i\otimes I$, $\sigma_1 \otimes
\sigma_i$, $\sigma_2 \otimes \sigma_i$,  $\sigma_3 \otimes
\sigma_i$\} for i=1, 2, 3. The density matrix is written as
\cite{9,10,11}
\begin{equation}
\rho=\frac{1}{4}(I + \sqrt{6} a \cdot J)
\end{equation}

The M matrix is given by:
\begin{widetext}
\begin{equation}
M=\frac{\sqrt{6}}{4}\left(
\begin{array}{llllllllllllllll}
 \frac{1}{\sqrt{6}} & 0 & 0 & 1 & 0 & 0 & 1 & 0 & 0 & 0 & 0 & 0 & 0 & 0 & 0 & 1 \\
 0 & 1 & -i & 0 & 0 & 0 & 0 & 0 & 0 & 0 & 0 & 0 & 0 & 1 & -i & 0 \\
 0 & 0 & 0 & 0 & 1 & -i & 0 & 0 & 0 & 1 & 0 & 0 & -i & 0 & 0 & 0 \\
 0 & 0 & 0 & 0 & 0 & 0 & 0 & 1 & -i & 0 & -i & -1 & 0 & 0 & 0 & 0 \\
 0 & 1 & i & 0 & 0 & 0 & 0 & 0 & 0 & 0 & 0 & 0 & 0 & 1 & i & 0 \\
 \frac{1}{\sqrt{6}} & 0 & 0 & -1 & 0 & 0 & 1 & 0 & 0 & 0 & 0 & 0 & 0 & 0 & 0 & -1 \\
 0 & 0 & 0 & 0 & 0 & 0 & 0 & 1 & i & 0 & -i & 1 & 0 & 0 & 0 & 0 \\
 0 & 0 & 0 & 0 & 1 & -i & 0 & 0 & 0 & -1 & 0 & 0 & i & 0 & 0 & 0 \\
 0 & 0 & 0 & 0 & 1 & i & 0 & 0 & 0 & 1 & 0 & 0 & i & 0 & 0 & 0 \\
 0 & 0 & 0 & 0 & 0 & 0 & 0 & 1 & -i & 0 & i & 1 & 0 & 0 & 0 & 0 \\
 \frac{1}{\sqrt{6}} & 0 & 0 & 1 & 0 & 0 & -1 & 0 & 0 & 0 & 0 & 0 & 0 & 0 & 0 & -1 \\
 0 & 1 & -i & 0 & 0 & 0 & 0 & 0 & 0 & 0 & 0 & 0 & 0 & -1 & i & 0 \\
 0 & 0 & 0 & 0 & 0 & 0 & 0 & 1 & i & 0 & i & -1 & 0 & 0 & 0 & 0 \\
 0 & 0 & 0 & 0 & 1 & i & 0 & 0 & 0 & -1 & 0 & 0 & -i & 0 & 0 & 0 \\
 0 & 1 & i & 0 & 0 & 0 & 0 & 0 & 0 & 0 & 0 & 0 & 0 & -1 & -i & 0 \\
 \frac{1}{\sqrt{6}} & 0 & 0 & -1 & 0 & 0 & -1 & 0 & 0 & 0 & 0 & 0 & 0 & 0 & 0 & 1
\end{array}
\right)
\end{equation}
\end{widetext}

Let's take the T matrix as in the previous case
\begin{equation}
T=diag\{1, \nu_1 ... \nu_{15}\}.
\end{equation}

B matrix is given by
\begin{widetext}
\begin{align}
 B = \frac{1}{4} & \tiny \left(
\begin{array}{ccccccc}
 1+\nu _3+\nu _6+\nu _{15} & 0 & 0 & 0 & 0 & \nu _1+\nu _2+\nu _{13}+\nu _{14} & 0 \\
 0 & 1-\nu _3+\nu _6-\nu _{15} & 0 & 0 & \nu _1-\nu _2+\nu _{13}-\nu _{14} & 0 & 0 \\
 0 & 0 & 1+\nu _3-\nu _6-\nu _{15} & 0 & 0 & 0 & 0 \\
 0 & 0 & 0 & 1-\nu _3-\nu _6+\nu _{15} & 0 & 0 & \nu _1-\nu _2-\nu _{13}+\nu _{14} \\
 0 & \nu _1-\nu _2+\nu _{13}-\nu _{14} & 0 & 0 & 1-\nu _3+\nu _6-\nu _{15} & 0 & 0 \\
 \nu _1+\nu _2+\nu _{13}+\nu _{14} & 0 & 0 & 0 & 0 & 1+\nu _3+\nu _6+\nu _{15} & 0 \\
 0 & 0 & 0 & \nu _1-\nu _2-\nu _{13}+\nu _{14} & 0 & 0 & 1-\nu _3-\nu _6+\nu _{15} \\
 0 & 0 & \nu _1+\nu _2-\nu _{13}-\nu _{14} & 0 & 0 & 0 & 0 \\
 0 & 0 & \nu _4-\nu _5+\nu _9-\nu _{12} & 0 & 0 & 0 & 0 \\
 0 & 0 & 0 & \nu _4-\nu _5-\nu _9+\nu _{12} & 0 & 0 & \nu _7-\nu _8-\nu _{10}+\nu _{11} \\
 \nu _4+\nu _5+\nu _9+\nu _{12} & 0 & 0 & 0 & 0 & \nu _7+\nu _8+\nu _{10}+\nu _{11} & 0 \\
 0 & \nu _4+\nu _5-\nu _9-\nu _{12} & 0 & 0 & \nu _7-\nu _8+\nu _{10}-\nu _{11} & 0 & 0 \\
 0 & 0 & 0 & \nu _7-\nu _8-\nu _{10}+\nu _{11} & 0 & 0 & \nu _4-\nu _5-\nu _9+\nu _{12} \\
 0 & 0 & \nu _7+\nu _8-\nu _{10}-\nu _{11} & 0 & 0 & 0 & 0 \\
 0 & \nu _7-\nu _8+\nu _{10}-\nu _{11} & 0 & 0 & \nu _4+\nu _5-\nu _9-\nu _{12} & 0 & 0 \\
 \nu _7+\nu _8+\nu _{10}+\nu _{11} & 0 & 0 & 0 & 0 & \nu _4+\nu _5+\nu _9+\nu _{12} & 0
\end{array}
\right| \nonumber \\
& \tiny \left| \begin{array}{ccccccc}
 0 & 0 & 0 & \nu _4+\nu _5+\nu _9+\nu _{12} & 0 & 0 & 0 \\
 0 & 0 & 0 & 0 & \nu _4+\nu _5-\nu _9-\nu _{12} & 0 & 0 \\
 \nu _1+\nu _2-\nu _{13}-\nu _{14} & \nu _4-\nu _5+\nu _9-\nu _{12} & 0 & 0 & 0 & 0 & \nu _7+\nu _8-\nu _{10}-\nu _{11} \\
 0 & 0 & \nu _4-\nu _5-\nu _9+\nu _{12} & 0 & 0 & \nu _7-\nu _8-\nu _{10}+\nu _{11} & 0 \\
 0 & 0 & 0 & 0 & \nu _7-\nu _8+\nu _{10}-\nu _{11} & 0 & 0 \\
 0 & 0 & 0 & \nu _7+\nu _8+\nu _{10}+\nu _{11} & 0 & 0 & 0 \\
 0 & 0 & \nu _7-\nu _8-\nu _{10}+\nu _{11} & 0 & 0 & \nu _4-\nu _5-\nu _9+\nu _{12} & 0 \\
 1+\nu _3-\nu _6-\nu _{15} & \nu _7+\nu _8-\nu _{10}-\nu _{11} & 0 & 0 & 0 & 0 & \nu _4-\nu _5+\nu _9-\nu _{12} \\
 \nu _7+\nu _8-\nu _{10}-\nu _{11} & 1+\nu _3-\nu _6-\nu _{15} & 0 & 0 & 0 & 0 & \nu _1+\nu _2-\nu _{13}-\nu _{14} \\
 0 & 0 & 1-\nu _3-\nu _6+\nu _{15} & 0 & 0 & \nu _1-\nu _2-\nu _{13}+\nu _{14} & 0 \\
 0 & 0 & 0 & 1+\nu _3+\nu _6+\nu _{15} & 0 & 0 & 0 \\
 0 & 0 & 0 & 0 & 1-\nu _3+\nu _6-\nu _{15} & 0 & 0 \\
 0 & 0 & \nu _1-\nu _2-\nu _{13}+\nu _{14} & 0 & 0 & 1-\nu _3-\nu _6+\nu _{15} & 0 \\
 \nu _4-\nu _5+\nu _9-\nu _{12} & \nu _1+\nu _2-\nu _{13}-\nu _{14} & 0 & 0 & 0 & 0 & 1+\nu _3-\nu _6-\nu _{15} \\
 0 & 0 & 0 & 0 & \nu _1-\nu _2+\nu _{13}-\nu _{14} & 0 & 0 \\
 0 & 0 & 0 & \nu _1+\nu _2+\nu _{13}+\nu _{14} & 0 & 0 & 0
\end{array}
\right| \nonumber \\
& \tiny  \left|
\begin{array}{cc}
 0 & \nu _7+\nu _8+\nu _{10}+\nu _{11} \\
 \nu _7-\nu _8+\nu _{10}-\nu _{11} & 0 \\
 0 & 0 \\
 0 & 0 \\
 \nu _4+\nu _5-\nu _9-\nu _{12} & 0 \\
 0 & \nu _4+\nu _5+\nu _9+\nu _{12} \\
 0 & 0 \\
 0 & 0 \\
 0 & 0 \\
 0 & 0 \\
 0 & \nu _1+\nu _2+\nu _{13}+\nu _{14} \\
 \nu _1-\nu _2+\nu _{13}-\nu _{14} & 0 \\
 0 & 0 \\
 0 & 0 \\
 1-\nu _3+\nu _6-\nu _{15} & 0 \\
 0 & 1+\nu _3+\nu _6+\nu _{15}
\end{array}
\right).
\end{align}
\end{widetext}

The eigenvalues of B are
\begin{widetext}
\begin{eqnarray}
\{\lambda_i\}=\{\frac{1}{4} \left(1+\nu _1+\nu _2+\nu _3+\nu
_4-\nu _5-\nu _6+\nu _7+\nu _8+\nu _9-\nu _{10}-\nu _{11}-\nu
_{12}-\nu
_{13}-\nu _{14}-\nu _{15}\right), \nonumber\\
 \frac{1}{4} \left(1+\nu
_1+\nu _2+\nu _3-\nu _4+\nu _5-\nu _6-\nu _7-\nu _8-\nu _9+\nu
_{10}+\nu _{11}+\nu _{12}-\nu _{13}-\nu _{14}-\nu _{15}\right),\nonumber\\
\frac{1}{4} \left(1+\nu _1-\nu _2-\nu _3+\nu _4+\nu _5+\nu _6+\nu
_7-\nu _8-\nu _9+\nu _{10}-\nu _{11}-\nu _{12}+\nu _{13}-\nu
_{14}-\nu _{15}\right),\nonumber\\
\frac{1}{4} \left(1+\nu _1-\nu _2-\nu _3-\nu _4-\nu _5+\nu _6-\nu
_7+\nu _8+\nu _9-\nu _{10}+\nu _{11}+\nu _{12}+\nu _{13}-\nu
_{14}-\nu _{15}\right),\nonumber\\
\frac{1}{4} \left(1-\nu _1+\nu _2-\nu _3+\nu _4+\nu _5+\nu _6-\nu
_7+\nu _8-\nu _9-\nu _{10}+\nu _{11}-\nu _{12}-\nu _{13}+\nu
_{14}-\nu _{15}\right),\nonumber\\
\frac{1}{4} \left(1-\nu _1+\nu _2-\nu _3-\nu _4-\nu _5+\nu _6+\nu
_7-\nu _8+\nu _9+\nu _{10}-\nu _{11}+\nu _{12}-\nu _{13}+\nu
_{14}-\nu _{15}\right),\nonumber\\
\frac{1}{4} \left(1-\nu _1-\nu _2+\nu _3+\nu _4-\nu _5-\nu _6-\nu
_7-\nu _8+\nu _9+\nu _{10}+\nu _{11}-\nu _{12}+\nu _{13}+\nu
_{14}-\nu _{15}\right),\nonumber\\
\frac{1}{4} \left(1-\nu _1-\nu _2+\nu _3-\nu _4+\nu _5-\nu _6+\nu
_7+\nu _8-\nu _9-\nu _{10}-\nu _{11}+\nu _{12}+\nu _{13}+\nu
_{14}-\nu _{15}\right),\nonumber\\
\frac{1}{4} \left(1-\nu _1-\nu _2+\nu _3-\nu _4-\nu _5+\nu _6+\nu
_7+\nu _8-\nu _9+\nu _{10}+\nu _{11}-\nu _{12}-\nu _{13}-\nu
_{14}+\nu _{15}\right),\nonumber\\
\frac{1}{4} \left(1-\nu _1-\nu _2+\nu _3+\nu _4+\nu _5+\nu _6-\nu
_7-\nu _8+\nu _9-\nu _{10}-\nu _{11}+\nu _{12}-\nu _{13}-\nu
_{14}+\nu _{15}\right),\nonumber\\
\frac{1}{4} \left(1-\nu _1+\nu _2-\nu _3-\nu _4+\nu _5-\nu _6+\nu
_7-\nu _8+\nu _9-\nu _{10}+\nu _{11}-\nu _{12}+\nu _{13}-\nu
_{14}+\nu _{15}\right),\nonumber\\
\frac{1}{4} \left(1-\nu _1+\nu _2-\nu _3+\nu _4-\nu _5-\nu _6-\nu
_7+\nu _8-\nu _9+\nu _{10}-\nu _{11}+\nu _{12}+\nu _{13}-\nu
_{14}+\nu _{15}\right),\nonumber\\
\frac{1}{4} \left(1+\nu _1-\nu _2-\nu _3-\nu _4+\nu _5-\nu _6-\nu
_7+\nu _8+\nu _9+\nu _{10}-\nu _{11}-\nu _{12}-\nu _{13}+\nu
_{14}+\nu _{15}\right),\nonumber\\
\frac{1}{4} \left(1+\nu _1-\nu _2-\nu _3+\nu _4-\nu _5-\nu _6+\nu
_7-\nu _8-\nu _9-\nu _{10}+\nu _{11}+\nu _{12}-\nu _{13}+\nu
_{14}+\nu _{15}\right),\nonumber\\
\frac{1}{4} \left(1+\nu _1+\nu _2+\nu _3-\nu _4-\nu _5+\nu _6-\nu
_7-\nu _8-\nu _9-\nu _{10}-\nu _{11}-\nu _{12}+\nu _{13}+\nu
_{14}+\nu _{15}\right),\nonumber\\
\frac{1}{4} \left(1+\nu _1+\nu _2+\nu _3+\nu _4+\nu _5+\nu _6+\nu
_7+\nu _8+\nu _9+\nu _{10}+\nu _{11}+\nu _{12}+\nu _{13}+\nu
_{14}+\nu _{15}\right)\}.
\end{eqnarray}

\end{widetext}

 Since all of the eigenvalues are hyperplanes, CP region is a simplex. Also,
 the vertices of this polyhedron correspond to the maps $1 \rho 1$,
 $J_i \rho J_i$, identical to the qubit maps. The extremal map - $J_i \rho J_i$ is just
 a rotation of the Bloch ball about the $J_i$ axis. The general
 depolarizing map can be written as a convex sum of these sixteen extremal maps.

\section{Discussion} We found that the general depolarizing map in N=2 and N=4 cases
has a simple geometric structure, while for N=3 it is much more
complicated. We conjecture that for any $2^n$ dimensional case the
CP region will be a simplex, with vertices corresponding to a
unitary rotation of the Block ball. It would be interesting to
explore the extent to which the N=2 and N=4 cases are similar for
other maps.

  We would also like to point out that our method of calculating
  B, and thus finding out the region where a map is CP
  is general and is not restricted to depolarizing maps. The most
  general map (Affine map) can be
  easily obtained if we let T also contain translations (see
  \cite{12}). In that case T becomes
  \begin{equation}
T=\left(%
\begin{array}{cc}
  1 & 0\\
  \overrightarrow{t} & T' \\
\end{array}%
\right)
\end{equation}
where T' contains only the unital part.

 K. D. would like to thank Cesar Rodriguez and Kavan Modi for
 helpful discussion and suggestions.

\end{document}